\documentclass[11pt]{article}
\usepackage{epsfig,cite}
\usepackage{amsmath}
\usepackage{amssymb}
\usepackage{amsfonts}
\usepackage{latexsym}
\usepackage{wasysym}
\usepackage{bm}
\catcode`@=11 

\newcommand{\beq}{\begin{equation}}
\newcommand{\eeq}{\end{equation}}
\newcommand{\bea}{\begin{eqnarray}}
\newcommand{\eea}{\end{eqnarray}}
\newcommand{\bdm}{\begin{displaymath}}
\newcommand{\edm}{\end{displaymath}}
\newcommand{\eps}{\varepsilon}

\newcommand{\kt}{\textrm{\textbf{k}}}

\newcommand{\qt}{\textrm{\textbf{q}}}

\def\quarter{\smallfrac{1}{4}}
\def\as{\alpha_s}

\def\MS{\hbox{$\overline{\rm MS}$}}

\def\bare{{(0)}}
\def\qb{\bar{q}}
\def\Qb{\bar{Q}}
\def\QN{\langle Q\rangle}
\def\QNbare{\langle Q^\bare\rangle}
\def\NN{\langle N\rangle}
\def\NNbare{\langle N^\bare\rangle}

\def\QQb{\{Q\bar{Q}\}}
\def\QQbN{\langle Q\bar{Q}\rangle}
\def\Nsq{\{N^2\}}
\def\NsqN{\langle N^2\rangle}
\def\Vsq{\{V^2\}}
\def\VsqN{\langle V^2\rangle}

\def\eqn#1{(\ref{#1})}
\def\m{{\cal M}}

\def\ord{{\cal O}}
\def\eps{\varepsilon}

\def\smallfrac#1#2{\hbox{${{#1}\over {#2}}$}}

\def \msb{\overline{\textrm{MS}}}

\newsavebox\tmpfig

\begin{document}

\pagestyle{empty}

\begin{flushright}

MAN/HEP/2008/46 \\
Edinburgh-2008/47\\ 
\end{flushright}

\begin{center}
\vspace*{2.5cm}
{\Large \bf High Energy Resummation of Drell-Yan Processes}
 \\
\vspace*{2.5cm}
Simone Marzani$^{a}$, Richard D.~Ball$^{b}$,
\\
\vspace{0.6cm}  {\it
{}$^a$ School of Physics and Astronomy, University of Manchester,\\
Oxford Road, Manchester M13 9PL, England, UK\\ \medskip
{}$^b$ School of Physics, University of Edinburgh,\\
Mayfield Rd, Edinburgh EH9 3JZ, Scotland, UK\\
}
\vspace*{2.5cm}

{\bf Abstract}
\end{center}

\noindent
We present a computation of the inclusive Drell-Yan production cross-section
in perturbative QCD to all orders in the limit of high partonic
centre--of--mass energy. We compare our results to the fixed order NLO
and NNLO results in $\msb$ scheme, and provide predictions at N$^3$LO
and beyond. Our expressions may be used to obtain fully resummed results
for the inclusive cross-section.

\vspace*{1cm}

\vfill
\noindent

\begin{flushleft} December 2008 \end{flushleft}
\eject

\setcounter{page}{1} \pagestyle{plain}

\section{Introduction}

Accurate perturbative calculations of benchmark inclusive cross-sections
are an essential component of the LHC discovery programme. The most
important of these benchmark processes are the Drell-Yan processes:
production of $\mu^+$-$\mu^-$ pairs, and the closely related vector
boson production processes. Currently these cross-sections are known at NNLO
in perturbative QCD \cite{Altarelli:1978id,Matsuura:1990ba,
Hamberg:1990np,Blumlein:2005im}. The resummation of threshold logarithms
is known up to N$^3$LL \cite{Catani:1989ne,Catani:2003zt,moch,magnea}.

When the invariant mass $Q$ of the particles produced in the final state is
well below the centre-of-mass energy, the typical values of $x$ of the
colliding partons may be rather small: $x_1x_2 = Q^2/S\ll 1$. In
fact this is true of most LHC processes: only when producing very
massive states close to threshold do both partons carry a large
fraction of the incoming longitudinal momentum. Whenever $x$ is
small, logarithms of $x$ may spoil the perturbation series. Thus
accurate calculations require the computation of the coefficients
of these logarithms, and if they are large
it may be necessary to resum them.

The resummation of small-$x$ logarithms in the perturbative evolution
of parton distribution functions at NLL is by now well understood
(see for example ref.\cite{Altarelli:2008xp} for a recent review).
The general procedure for resumming hard cross-sections at
the same order through $k_T$-factorization is known \cite{CCH-photoprod,ch},
and its implementation when the coupling runs understood \cite{ball}.
Calculations have been performed for photoproduction processes
\cite{CCH-photoprod,ball}, deep inelastic processes \cite{ch,Altarelli:2008aj},
hadroproduction of heavy quarks \cite{CCH-photoprod,ellis-hq,camici-hq,ball},
and gluonic Higgs production both in the pointlike limit
\cite{Hautmann:2002tu}, and for finite top mass
\cite{Marzani:2008az,Marzani:2008ih}. However for Drell-Yan and vector
boson production the resummed hard cross-section has yet to be computed.
It is the purpose of this paper to perform this calculation, determining
the coefficients of the leading high energy singularities of the Drell-Yan
coefficient function to all orders in perturbation theory. This will enable
the accurate evaluation of small-$x$ corrections to these benchmark processes,
soon to be measured at LHC.

\section{The Drell-Yan cross-section}

We wish to study the high energy behaviour of the Drell-Yan cross-section.
We consider $n_f$ quarks $q_i$ with electric
charge $e_i$.
The cross-section for the production of a lepton pair via
an off-shell photon with squared momentum $q^2=Q^2$ can then be written as:
\begin{equation}\label{xseccon}
 \sigma(\tau_h, Q^2) = \sigma_0(Q^2)\sum_{a,b=q_i,\bar{q_j},g}
\int_\rho^1 \! {dx_1\over x_1}\int_\rho^1 \! {dx_2\over x_2}
D_{ab}(\smallfrac{\tau_h}{x_1x_2};\as(Q^2))F_a(x_1,Q^2)F_b(x_2,Q^2)\,,
\end{equation}
where $\tau_h= Q^2/S$, and $F_a(x,Q^2)=xf_a(x,Q^2)$ is the integrated
parton density for parton $a$,
the indices $a$ and $b$ running over the different initial
partons \mbox{$q_i,\, \bar{q_j},\,g$}.
The LO partonic cross-section is simply
\begin{equation} \label{dylo}
\sigma_0(Q^2)= \frac{\alpha}{Q^2}\frac{4 \pi}{3N_c}\langle e^2 \rangle\,,
\end{equation}
where $\langle e^2 \rangle\equiv \smallfrac{1}{n_f}\sum_i e_i^2$:
the dimensionless coefficient functions $D_{ab}$ then contain all the QCD
radiative corrections.

\begin{figure}[t!]
\begin{center}
\includegraphics[width= 12cm]{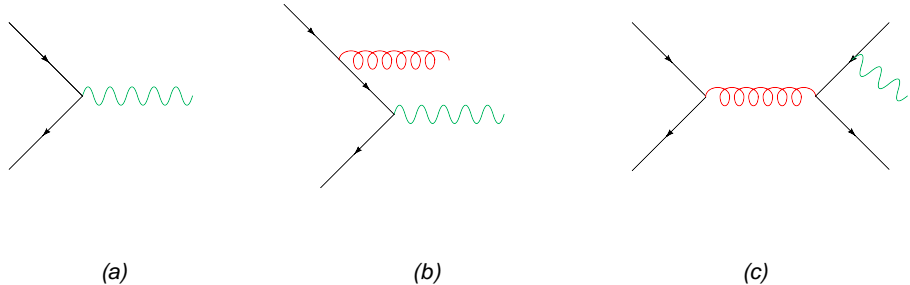}
\caption{(a) LO and (b) NLO diagrams contributing to $D_{q\qb}$, and
(c) a NNLO diagram contributing to the flavour disconnected
piece $\tilde D_{q\qb}$.
Both quark and antiquark must carry the same flavour. \label{dyqq}}
\end{center}
\end{figure}

In what follows we will employ a Mellin transform in $\tau_h$ to undo the
convolutions in eqn.~(\ref{xseccon}): defining
\begin{equation}\label{meldef}
\sigma(N, Q^2) = \int_0^1 \!d\tau_h \tau_h^{N-1} \sigma(\tau_h,Q^2),
\end{equation}
the factorization eqn.~(\ref{xseccon}) becomes simply
\begin{equation}\label{xsecmel}
 \sigma(N, Q^2) = \sigma_0(Q^2)\sum_{a,b=q_i,\bar{q_j},g}
D_{ab}(N;\as(Q^2))F_a(N,Q^2)F_b(N,Q^2)\,.
\end{equation}
Clearly $D_{ab}=D_{ba}$, and thus since the strong interaction is CP even
$D_{q_iq_j}=D_{\qb_i\qb_j}$, $D_{q_i\qb_j}=D_{\qb_iq_j}$ and
$D_{q_ig}=D_{\qb_i g}$.
Thus if we define singlet and nonsinglet quark (plus antiquark) distributions
and quark-antiquark luminosities as
\bea\label{qSNS}
Q &\equiv& \sum_i (Q_i+\bar{Q}_i), \qquad
\QN \equiv \smallfrac{1}{\langle e^2\rangle}\sum_i e_i^2 (Q_i+\bar{Q}_i)\\
\QQb &\equiv& \sum_i Q_i\bar{Q}_i\qquad
\QQbN \equiv \smallfrac{1}{\langle e^2\rangle}\sum_i e_i^2 Q_i\bar{Q}_i,
\nonumber
\eea
where $Q_i=xq_i$ etc, then we can decompose the factorization eqn.~(\ref{xsecmel}) as
\bea
\sigma&=&\sigma_0\big[D_{q\qb}\QQbN
+ D_{qq}\QN Q
+ D_{qg}\QN G\nonumber\\
&&\qquad\qquad + n_f\tilde{D}_{q\qb}\QQb
+n_f\tilde D_{qq} QQ +2 n_f\tilde D_{qg}Q  G
+n_f\tilde{D}_{gg}GG\big].\label{xsecmelD}
\eea
Inevitably this flavour decomposition is rather more complicated
than in DIS (see ref.\cite{ch}) because here one must consider the
flavours of three partons: the two incoming ones and the one from which
the virtual photon is emitted.

\begin{figure}[t!]
\begin{center}
\includegraphics[width= 8cm]{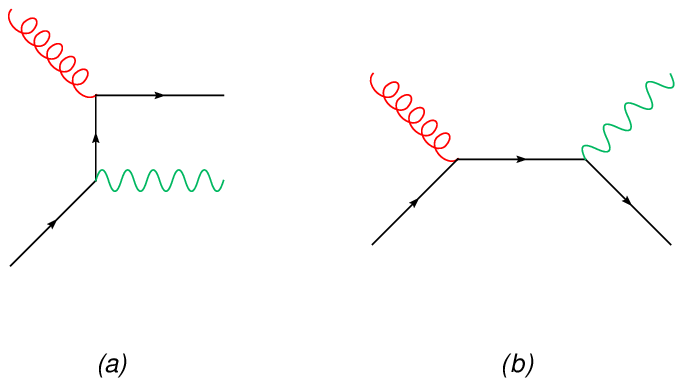}
\caption{NLO diagrams contributing to $D_{qg}$.\label{dyqg}}
\end{center}
\begin{center}
\includegraphics[width= 12cm]{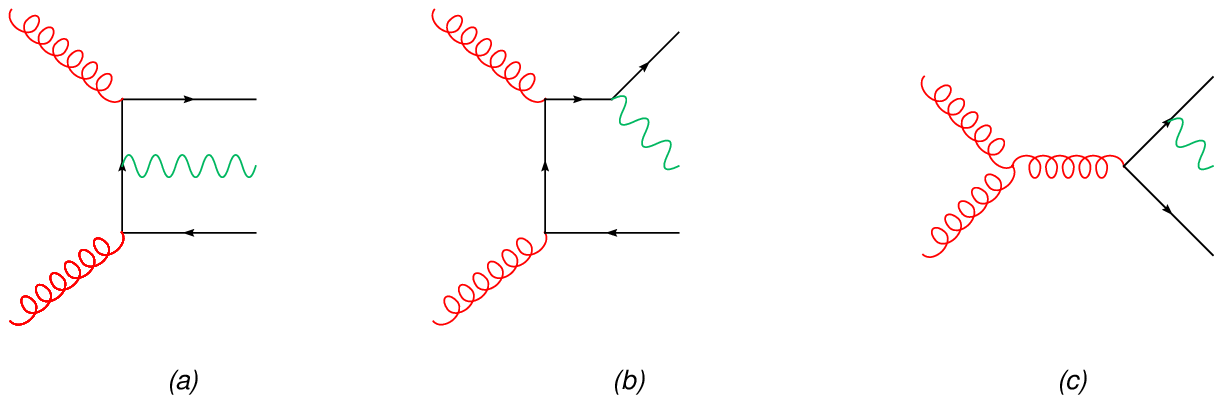}
\end{center}
\caption{NNLO diagrams contributing to $\tilde{D}_{gg}$ at NNLO.
\label{dygg}}
\end{figure}

\begin{figure}[t!]
\begin{center}
\includegraphics[width= 12cm]{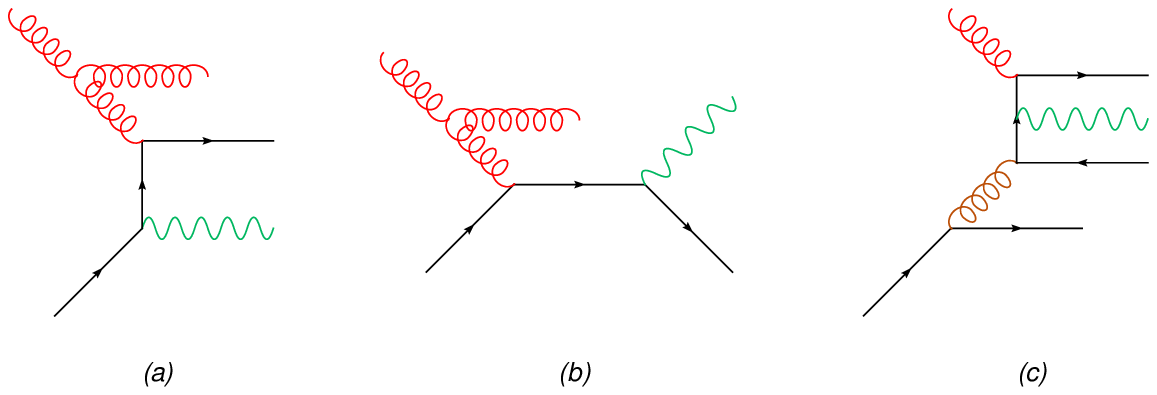}
\end{center}
\caption{(a) and (b) are NNLO diagrams contributing to $D_{qg}$, which
introduce high energy logarithms, while (c) is a N$^3$LO diagram
contributing to the flavour disconnected piece $\tilde D_{qg}$.
\label{dyqgg}}
\begin{center}
\includegraphics[width= 12cm]{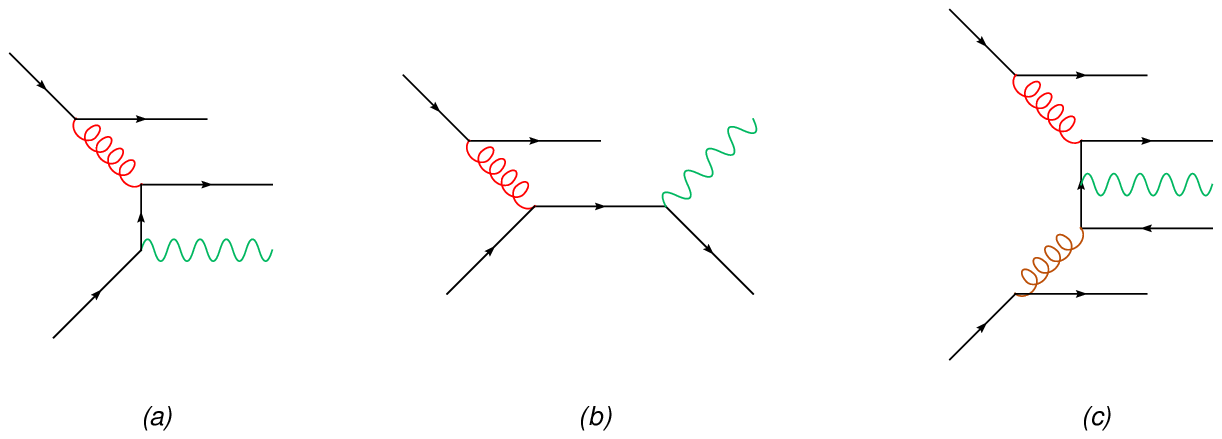}
\end{center}
\caption{(a) and (b) are NNLO diagrams contributing to $D_{qq}$, while
(c) is an N$^4$LO diagram contributing to the flavour disconnected piece
$\tilde D_{qq}$.\label{dyqqg}}
\end{figure}

We now consider the behaviour of the perturbation series for
the coefficients $D_{ab}(N;\as(Q^2))$. At LO the only
process is $q \bar{q} \to \gamma^*$ and thus $D_{q\qb}=1$ (fig.\ref{dyqq}a),
while all the rest are zero.
At NLO the channel $q g \to q\gamma^*$ opens: $D_{qg}=\ord(\as)$
(fig.\ref{dyqg}). Processes with two gluons, two quarks, or two antiquarks
in the initial state can only occur at NNLO:
$\tilde D_{gg}=\ord(\as^2)$ (fig.\ref{dygg}), and $D_{qq}=\ord(\as^2)$
(fig.\ref{dyqqg}a and \ref{dyqqg}b). At this order there is also a
``flavour disconnected'' process contributing to $q \bar{q} \to \gamma^*$:
$\tilde D_{q\qb}=\ord(\as^2)$ (fig.\ref{dyqq}c). Flavour disconnected
processes contribute also to $q g \to X\gamma^*$
($\tilde D_{qg}=\ord(\as^3)$, fig.\ref{dyqgg}c) and
$q q \to X\gamma^*$ ($\tilde D_{qq}=\ord(\as^4)$, fig.\ref{dyqqg}c): note
that these each have a $gg\to q\qb \gamma^*$ subprocess. It is
easy to see that all these flavour disconnected pieces are two powers of
$\as$ down compared to the corresponding flavour connected piece, and
that they must be proportional to $n_f$.

Now consider the behaviour of the coefficient functions at high energy,
that is as $N\to 0$. Only some of the higher order corrections contain
high energy
logarithms, and these are the ones we wish to isolate and compute.
To see the how these logarithms arise, and thus deduce the
general pattern, note that at LO the $gg$
anomalous is singular, $\gamma_{gg}\sim \smallfrac{\as}{N}$,
while the $qq$ anomalous
dimension is not, $\gamma_{qq}\sim \as$. Thus the NLO corrections to
$D_{q\qb}$ due to gluon emission from a quark line (fig.\ref{dyqq}b)
are $\ord(\as)$: further gluon emissions from quark lines give
contributions of $\ord(\as^k)$. Similarly the flavour disconnected piece
$\tilde D_{q\qb}$ has no high energy logarithms. In
fact both the terms in eqn.~(\ref{xsecmelD}) in which an
incoming quark and antiquark annihilate to produce the vector boson are
completely regular as $N\to 0$, in the same way
that the nonsinglet contribution to DIS structure functions is regular:
\beq\label{heNS}
D_{q\qb} = 1 +
C_F(\smallfrac{4}{3}\pi^2-\smallfrac{7}{2})\smallfrac{\as}{2\pi}
+\ord(\as^2),\qquad \tilde{D}_{q\qb} = \ord(\as^2).
\eeq

However the NNLO contribution to $D_{qg}$ can 
be singular, due to gluon emission from the incoming gluon line 
(fig.\ref{dyqgg}a and \ref{dyqgg}b), 
and is thus $\ord(\as\smallfrac{\as}{N})$. Further gluon 
emissions from the incoming gluon give extra powers of $\smallfrac{\as}{N}$, 
while gluon emissions from a quark line just give an extra $\as$, so the most 
singular contributions are $\ord(\as(\smallfrac{\as}{N})^k)$, with $k\geq 0$, 
i.e. NLL$x$. Similarly the N$^3$LO contribution to $D_{gg}$ is 
$\ord(\as^2\smallfrac{\as}{N})$, due to gluon emission from one 
of the incoming gluons: in general the most singular contributions here are 
$\ord(\as^2(\smallfrac{\as}{N})^k)$, $k\geq 0$ 
i.e. NNLL$x$. Finally consider $D_{qq}$: at
NNLO besides gluon emission from quark lines, one of the initial quarks 
can turn into a gluon, the rest of the process then proceeding as for $D_{qg}$ 
(fig.\ref{dyqqg}a and \ref{dyqqg}b). 
Since $\gamma_{gq}\sim \smallfrac{C_F}{C_A}\smallfrac{\as}{N}$, these 
graphs are also singular, $\ord(\as\smallfrac{\as}{N})$. 
Further emissions from 
the gluon line then give contributions of $\ord(\as(\smallfrac{\as}{N})^k)$, 
$k\geq 1$, and are thus again NLL$x$.
In fact these graphs are at high energy identical to the corresponding 
graphs for $D_{qg}$, save for the extra factor of $\smallfrac{C_F}{C_A}$ 
due to the incoming quark rather than incoming gluon, so there is a 
colour-charge relation\cite{ch} between the leading singularities 
of $D_{qq}$ and those of $D_{qg}$. 
Similar considerations apply to the disconnected contributions 
$\tilde D_{qq}$ and $\tilde D_{qg}$: the most singular contributions to 
$\tilde{D}_{qg}$ are $\ord(\as^2(\smallfrac{\as}{N})^k)$, $k\geq 1$, 
and related through a factor of $\smallfrac{C_F}{C_A}$ to those of $D_{gg}$ 
(see fig. \ref{dyqgg}c), while those of $\tilde{D}_{qq}$ are 
$\ord(\as^2(\smallfrac{\as}{N})^k)$, $k\geq 2$, and related through 
a factor of $(\smallfrac{C_F}{C_A})^2$ to those of $D_{gg}$ 
(see fig. \ref{dyqqg}c).

Summarising, in the high energy limit the leading singularities of the
coefficient functions have the schematic structure
\bea\label{heS}
D_{qq}(N,\as)&=& \smallfrac{C_F}{C_A}
\sum_{k=1}^\infty \ord(\as(\smallfrac{\as}{N})^k)\,, \nonumber \\
D_{qg}(N,\as)&=& \ord(\as)+\sum_{k=1}^\infty \ord(\as(\smallfrac{\as}{N})^k)\,,
\nonumber \\
\tilde{D}_{qq}(N,\as)&=& (\smallfrac{C_F}{C_A})^2
\sum_{k=2}^\infty \ord(\as^2(\smallfrac{\as}{N})^k)\,,  \\
\tilde{D}_{qg}(N,\as)&=& \smallfrac{C_F}{C_A}
\sum_{k=1}^\infty \ord(\as^2(\smallfrac{\as}{N})^k)\,,\nonumber \\
\tilde{D}_{gg}(N,\as)&=& \ord(\as^2)
+ \sum_{k=1}^\infty \ord(\as^2(\smallfrac{\as}{N})^k)
\,.\nonumber
\eea
It follows that while the quark-quark and quark-gluon coefficient functions
are NLL$x$, the gluon-gluon coefficient function and the disconnected
contributions to quark-quark and quark-gluon are all NNLL$x$, and
thus need not be considered in a NLL$x$ calculation (though they may still
be numerically large, since the gluon distribution is rather larger
than the quark distribution).
Our strategy in the rest of this paper will thus be to use
\mbox{$k_T$-factorization} to compute the leading high energy
singularities of the $D_{qg}$ coefficient function to all order in
perturbation theory, and then deduce those of $D_{qq}$
using colour-charge relations. The high energy singularities are found
by considering the off-shell process
$g^* q\to\gamma^* q $
calculated from the graphs in fig.~\ref{dyqg} with the incoming gluon off-shell,
in just the same way as one determines the coefficient function $C_{2g}$
in DIS using the off-shell process
$\gamma^* g^* \to q \qb $~\cite{ch}. The computation is
complicated by the fact
that just as in the DIS calculation there are collinear singularities
as well as high energy singularities, so one has to take care to
factorize the collinear singularities before extracting the high energy ones.
In the next section we explain in some detail how this can be done.

The hadroproduction of vector bosons, either $W^{\pm}$ or $Z$, has a
very similar structure to Drell-Yan: in Mellin space the cross
section can be written as
\begin{equation}\label{xsecmelV}
 \sigma^V(N, m_V^2) = \sigma^V_0\sum_{a,b=q,\bar{q},g}
D^V_{ab}(N;\as(Q^2))F_a(N,Q^2)F_b(N,Q^2)\,.
\end{equation}
where now
\beq
 {\sigma}_0^{V} = \frac{\pi}{N_c}\sqrt{2}G_F
\eeq
and $G_F$ is the Fermi constant. The decomposition eqn.~(\ref{xsecmel})
is the same up to threshold effects, except that for $Z$-production $e_i^2$
in eqn.~(\ref{qSNS}) is replaced by $v_i^2+a_i^2$, where $v_i$, $a_i$
are the vector and axial couplings of the $Z$ bosons to the  quarks, while
for $W^\pm$ production
\beq\label{SNSW}
\QN=Q,\qquad \QQbN = \sum_{ij}|V_{ij}|^2Q_i\Qb_j,
\eeq
where $V_{ij}$ is the CKM matrix, $\sum_jV_{ij}V_{jk}^*=\delta_{ik}$,
so the disconnected pieces of the quark-quark and quark-gluon
coefficient functions may be simply added to the connected pieces.
The perturbative QCD corrections to
vector boson production are the same as the Drell-Yan ones at NLO, and
they only begin to differ at NNLO because of the diagram with two incoming
gluons and an internal quark loop, which gives an $\ord(\as^2)$ contribution
to the $Z$ boson cross-section
but vanishes in the case of a virtual photon or $W^{\pm}$.
The NLL$x$ singularities are thus determined by the off-shell processes
$g^* q\to W^{\pm}q'$ and $g^* q\to Zq$
at $\ord(\as)$ and are the same as those for Drell-Yan:
the NNLL$x$ singularities for $Z$ production would however
receive an extra contribution.

\section{High energy factorization}

The general structure of high energy singularities outlined in the previous
section arises from the $k_T$-factorization formalism \cite{CCH-photoprod},
and in particular its extension by Catani and Hautmann to deal with
situations in which the hard cross-section contains collinear
singularities which must be factorized consistently \cite{ch}. In
this section we explain how this works for Drell-Yan processes.

We saw in the previous section how the singular contributions to the
cross-section eqn.~(\ref{xsecmelD}) are of essentially two kinds:
those in which the electroweak boson is emitted directly from one
of the incoming quarks or antiquarks, and those (the `disconnected' pieces)
where it is not. While the former can only acquire high energy
logarithms through gluon emission from the second parton, the
former acquire high energy logarithms through gluon emission from both
partons. The dimensional regularised form of $k_T$-factorization for
the singular part of the Drell-Yan cross-section is thus (in dimensional
regularization)
\bea \label{xsecmelSkt}
 \sigma_s(N, Q^2) &=&\sigma_0(Q^2) \bigg\{
\int \frac{d^{2-2 \eps}\kt}{\pi\kt^2}\,
\Sigma_{qg}(N,\smallfrac{\kt^2}{Q^2};\as,\eps)
\langle Q^\bare(N,\mu,\eps)\rangle \nonumber\\
&&\qquad\qquad\qquad
\big({\cal F}_q^\bare(N,\smallfrac{\kt^2}{\mu^2};\as,\eps)
Q^\bare(N,\mu,\eps)
+{\cal F}_g^\bare(N,\smallfrac{\kt^2}{\mu^2};\as,\eps)
G^\bare(N,\mu,\eps)\big)
\nonumber\\
&&+ n_f\int \frac{d^{2-2 \eps}\kt_1}{\pi\kt_1^2}\,
\frac{d^{2-2 \eps}\kt_2}{\pi\kt_2^2}\,
\Sigma_{gg}(N,\smallfrac{\kt_1^2}{Q^2},\smallfrac{\kt_2^2}{Q^2};\as,\eps)
\nonumber\\
&&\qquad\qquad\qquad
\big({\cal F}_q^\bare(N,\smallfrac{\kt_1^2}{\mu^2};\as,\eps)
Q^\bare(N,\mu,\eps)
+{\cal F}_g^\bare(N,\smallfrac{\kt_1^2}{\mu^2};\as,\eps)
G^\bare(N,\mu,\eps)\big)
\nonumber\\
&&\qquad\qquad\qquad
\big({\cal F}_q^\bare(N,\smallfrac{\kt_2^2}{\mu^2};\as,\eps)
Q^\bare(N,\mu,\eps)
+{\cal F}_g^\bare(N,\smallfrac{\kt_2^2}{\mu^2};\as,\eps)
G^\bare(N,\mu,\eps)\big)\bigg\}, \nonumber \\
\eea
where $\Sigma_{qg}(N,\smallfrac{\kt^2}{Q^2};\as,\eps)$
is the partonic cross-section
for $qg\to \gamma^*q$ with the incoming gluon off-shell (given at
leading order by the diagrams fig.~\ref{dyqg}, and thus $\ord(\as)$),
$\Sigma_{gg}(N,\smallfrac{\kt_1^2}{Q^2},
\smallfrac{\kt_2^2}{Q^2};\as,\eps)$
is the partonic cross-section
for $gg\to \gamma^*q\qb$ with both the incoming gluons off-shell
(given at leading order by the diagrams fig.~\ref{dygg}, and
thus $\ord(\as^2)$),
${\cal F}_q^\bare$ and ${\cal F}_g^\bare$ are the
bare quark and gluon Green's functions, and $Q^\bare$ and
$G^\bare$ the bare integrated quark (plus antiquark) and
gluon densities.\footnote{Note that our notation is slightly
different to that in \cite{ch}, since in particular we
prefer to deal throughout with integrated parton densities.}
At LL$x$ the quark Green's function is
\begin{equation}\label{qgcc}
{\cal F}_q^\bare(N,\smallfrac{\kt^2}{\mu^2};\as,\eps)
=\smallfrac{C_F}{C_A}
\left[{\cal F}_g^\bare(N,\smallfrac{\kt^2}{\mu^2};\as,\eps)
-\pi\kt^2\delta^{(2-2 \eps)}(\kt) \right]\,,
\end{equation}
while the gluon Green's function is
\begin{equation}\label{ggf}
{\cal F}_g^\bare(N,\smallfrac{\kt^2}{\mu^2};\as,\eps)
= \gamma_{gg} R(\gamma_{gg})
\Gamma_{gg}(N;\as,\eps)\left(\smallfrac{\kt^2}{\mu^2} \right)^{\gamma_{gg}} \,.
\end{equation}
The gluon-gluon transition function (containing the collinear poles)
is given by
\begin{equation}\label{transgg}
 \Gamma_{gg}(N;\as,\eps)= \exp \left(\frac{1}{\eps}\int_0^{\as S_{\eps}}
\frac{d \alpha}{\alpha} \gamma_{gg}(N,\alpha)\right)\,,
\end{equation}
where $S_{\eps}={\rm exp }[-\eps(-\gamma_E+ \ln 4 \pi)]$ and $R(\gamma_{gg})$ is a scheme-dependent
normalization factor, regular as $\eps\to 0$: in
the \MS\ scheme
\begin{equation}\label{are}
R(\gamma_{gg})=\left\{{\Gamma(1-\gamma_{gg})\chi_0(\gamma_{gg})
\over \Gamma(1+\gamma_{gg})[-\gamma_{gg}\chi_0'(\gamma_{gg})]}\right\}^{1/2}
\exp\left\{\gamma_{gg}\psi(1)+\int_0^{\gamma_{gg}}
d\gamma {\psi'(1)-\psi'(1-\gamma)\over
\chi_0(\gamma)}\right\} \,.
\end{equation}
In the high energy limit
\begin{equation}\label{ggghe}
 \gamma_{gg}(N,\as) =
\gamma_s \left(\smallfrac{\as}{N}\right)
+ \ord \left(\as \left(\smallfrac{\as}{N}\right)^k \right)\,.
\end{equation}
where $\gamma_s\left(\smallfrac{\as}{N}\right)
=\smallfrac{\as}{N}+\ord\left(\left(\smallfrac{\as}{N}\right)^4\right)$ is the
naive dual to the BFKL kernel $\chi_0$.

We can use the the $k_T$-factorization
eqn.~(\ref{xsecmelSkt}) to extract formulae for the bare coefficients
in the expression eqn.~(\ref{xsecmelD}) at high energy in terms of the
off-shell partonic cross-sections $\Sigma_{qg}$ and $\Sigma_{gg}$. For instance,
comparing the coefficients of $\QNbare G^\bare$, we have
\bea \label{dimregkt}
D_{qg}^\bare(N,Q^2;\as,\eps)
&=& \int \frac{d^{2-2 \eps}\kt}{\pi\kt^2}\,
{\Sigma}_{qg}(N,\frac{\kt^2}{Q^2};\as,\eps)
{\cal F}_g^\bare(N,\smallfrac{\kt^2}{\mu^2};\as,\eps)\nonumber\\
&=& \int \frac{d^{2-2 \eps}\kt}{\pi\kt^2}\,
{\Sigma}_{qg}(N,\smallfrac{\kt^2}{Q^2};\as,\eps)
\left(\frac{\kt^2}{\mu^2} \right)^{\gamma_{gg}}
 \gamma_{gg}R(\gamma_{gg})
\Gamma_{gg}(N;\as,\eps),
\eea
where in the second line we substituted the gluon Green's
function (\ref{ggf}). Introducing $\xi = \kt^2/Q^2$, and defining
the impact factor
\beq\label{impfact}
h_{qg}(N,M;\as)\equiv
M^2\int_0^\infty\!d\xi\, \xi^{M-1}
{\Sigma}_{qg}(N,\xi;\as,0)
\eeq
we can write this as
\beq\label{dimregqgbare}
D_{qg}^\bare(N,Q^2;\as,\eps)
= \left(\smallfrac{Q^2}{\mu^2} \right)^{\gamma_{gg}}
h_{qg}(N,\gamma_{gg};\as)
\gamma_{gg}^{-1} R(\gamma_{gg})
\Gamma_{gg}(N;\as,\eps)+\ord(\as^2(\smallfrac{\as}{N})^k).
\eeq
Note that we can set $\eps=0$ in the impact factor, since collinear
singularities are now regularised by $M\neq 0$. The reason for the
extra factor of $M^2$ in eqn.~\eqn{impfact} will become clear shortly.

Similarly, comparing coefficients of $\QNbare Q^\bare$ we find
\bea\label{dimregqqbare}
D_{qq}^\bare(N,Q^2;\as,\eps)
&=& \smallfrac{C_F}{C_A}\Big[\left(\smallfrac{Q^2}{\mu^2} \right)^{\gamma_{gg}}h_{qg}(N,\gamma_{gg};\as)
\gamma_{gg} R(\gamma_{gg})
\Gamma_{gg}(N;\as,\eps)\nonumber\\
&&\qquad\qquad-{\Sigma}_{qg}(N,0;\as,\eps)\Big]+
\ord(\as^2(\smallfrac{\as}{N})^k),\nonumber\\
&=& \smallfrac{C_F}{C_A}\Big[D_{qg}^\bare(N,Q^2;\as,\eps)
-{\Sigma}_{qg}(N,0;\as,\eps)\Big]+
\ord(\as^2(\smallfrac{\as}{N})^k),
\eea
the second term (which is also singular) coming from the delta-function
subtraction in eqn.~(\ref{qgcc}). Similar expressions for the bare
coefficients $\tilde D_{qq}$, $\tilde D_{qg}$ and $\tilde D_{gg}$ may be
extracted in terms of double Mellin transforms of $\Sigma_{gg}$: we will
ignore these in what follows as they are $\ord(\as^2)$.

\section{Collinear factorization at high energy}

We now need to factorize the collinear singularities in order to absorb
them into renormalized parton distribution functions. In the case of a collinear safe process
such as heavy flavour hadroproduction
\cite{ellis-hq,camici-hq} this is relatively straightforward: all the $\eps$ poles are in the
Green's functions, and
in the \MS\ factorization scheme are absorbed
into the bare gluon distribution, i.e. $G=\Gamma_{gg }G^\bare$.
However when we consider a process in which the partonic cross-section
is not collinear safe,
such as DIS or DY, the factorization of the collinear
singularities is rather more involved \cite{ch}.

Firstly, we must separate out the valence and nonsinglet quark
distributions from the singlet, since the former factorize
multiplicatively while the latter mixes with the gluon. Defining
\beq\label{VNS}
V_i\equiv Q_i-\Qb_i,\qquad N_i\equiv Q_i+\Qb_i - \smallfrac{1}{n_f}Q
\eeq
so $\sum_i N_i = 0$, we can rewrite the various quantities (\ref{qSNS})
appearing in eqn.~(\ref{xsecmelD})
\bea\label{qSNSV}
\QN &=&\smallfrac{1}{\langle e^2\rangle}
\sum_i e_i^2 (N_i+\smallfrac{1}{n_f}Q)=\NN+Q,\nonumber\\
\QQb
&=& \quarter \sum_i\big((N_i+\smallfrac{1}{n_f}Q)^2-V_i^2\big)
= \quarter\Nsq - \quarter\Vsq + \smallfrac{1}{4n_f}Q^2,\\
\QQbN&=&\quarter\smallfrac{1}{\langle e^2\rangle}\sum_i e_i^2
\big((N_i+\smallfrac{1}{n_f}Q)^2-V_i^2\big)
= \quarter\NsqN - \quarter\VsqN +\smallfrac{1}{2n_f}\NN Q+
\smallfrac{1}{4n_f}Q^2.
\nonumber
\eea
When written in terms of $N$, $V$, $Q$ and $G$
the Drell-Yan cross-section eqn.~(\ref{xsecmelD})
separates into three distinct pieces: $\sigma = \sigma^{NS}
+\sigma^S+\sigma^{SS}$, where
\bea
\sigma^{NS}&=&\sigma_0\big[\quarter D_{q\qb}(\NsqN-\VsqN)
+ \quarter n_f\tilde{D}_{q\qb}(\Nsq-\Vsq)\big],\label{sigNS}\\
\sigma^{S}&=&\sigma_0\big[(D_{qq}+ \smallfrac{1}{2n_f}D_{q\qb})\NN Q
+D_{qg}\NN G\big],\label{sigS}\\
\sigma^{SS}&=&\sigma_0\big[(D_{qq}+
\smallfrac{1}{4n_f}D_{q\qb})QQ
+D_{qg}QG\nonumber\\
&&\qquad+
(n_f\tilde D_{qq}+\smallfrac{1}{4n_f}
\tilde{D}_{q\qb})QQ +2n_f\tilde D_{qg}QG
+n_f D_{gg}GG\big].\label{sigSS}
\eea

Now the factorization of collinear singularities for the valence and
nonsinglet distributions $V_i$ and $N_i$ is multiplicative:
\beq\label{factVN}
V_i(N,\as) = \Gamma^V_{qq}(N,\as,\eps)V_i^\bare(N,\as,\eps),
\qquad N_i(N,\as) = \Gamma^N_{qq}(N,\as,\eps)N_i^\bare(N,\as,\eps)
\eeq
so the nonsinglet cross-section $\sigma^{NS}$ factorizes multiplicatively.
Furthermore the transition functions $\Gamma^V_{qq}$ and $\Gamma^N_{qq}$
contain no high energy singularities: at high energy the both reduce to unity
up to corrections of $\ord(\as)$. However singlet quark $Q$ and gluon $G$ mix:
\beq \label{factQG}
Q = \Gamma_{qq}\, Q^\bare+2n_f\Gamma_{qg}\, G^\bare \qquad
G = \Gamma_{gq}\, Q^\bare+\Gamma_{gg}\, G^\bare\,,
\eeq
and the transition functions now contain high energy singularities.
These were determined to NLL$x$ in \cite{ch}:
for example
\begin{equation}\label{transqg}
 \Gamma_{qg}(N,\as,\eps)= \frac{1}{\eps}\int_0^{\as S_{\eps}}
\frac{d \alpha}{\alpha} \gamma_{qg}(N,\alpha)\Gamma_{gg}(N,\alpha,\eps)\,,
 \end{equation}
where $\Gamma_{gg}$ is given by eqn.~(\ref{transgg}), and thus $\Gamma_{qg}$
contains (singular) terms of $\ord(\as(\smallfrac{\as}{N})^k)$. $\Gamma_{gq}$
and $\Gamma_{qq}$ are given in terms of $\Gamma_{gg}$ and $\Gamma_{qg}$
by colour-charge relations:
\beq\label{transgq}
\Gamma_{gq}(N,\as,\eps) = \smallfrac{C_F}{C_A}
\left[\Gamma_{gg}(N,\as,\eps)-1\right]
\eeq
while writing $\Gamma_{qq}=1+2n_f\Gamma_{qq}^{PS}$,
\beq
\Gamma_{qq}^{PS}(N,\as,\eps) = \frac{1}{\eps}\int_0^{\as S_{\eps}}
\frac{d \alpha}{\alpha} [\gamma_{qq}+\gamma_{qg}\Gamma_{gq}]
= \smallfrac{C_F}{C_A}\left[\Gamma_{qg}(N,\as,\eps)- \smallfrac{\as}{\eps}
\gamma_{qg}^0(N)\right],\label{transqq}
\eeq
since at high energy
$\gamma_{qq}(N,\as) = \smallfrac{C_F}{C_A}[\gamma_{qg}(N,\as)
-\as\gamma_{qg}^0(0)]$, where $\gamma_{qg}(N,\as)
= \as\gamma_{qg}^0(N)+\ord(\as^2)$.

Thus if we consider the collinear factorization of the singlet cross-section
$\sigma^{S}$, substituting eqns.~(\ref{factVN},\ref{factQG}) into
eqn.~(\ref{sigS}), the gluon and quark contributions to
$\sigma^S$ are given respectively by
\bea
\sigma_0^{-1} \sigma^S_{g}&\equiv&
D_{qg}^\bare \NNbare G^\bare
= \Gamma_{qq}^N\left[(D_{qq}+\smallfrac{1}{2n_f}D_{q\qb})2n_f\Gamma_{qg}
+D_{qg}\Gamma_{gg} \right]\NNbare G^\bare\,,\label{sigSg}\\
\sigma_0^{-1} \sigma^S_{q}
&\equiv& (D_{qq}^\bare+ \smallfrac{1}{2n_f}D_{q\qb}^\bare)\NNbare
Q^\bare = \Gamma_{qq}^N\left[(D_{qq}+ \smallfrac{1}{2n_f}D_{q\qb})\Gamma_{qq}
+D_{qg}\Gamma_{gq}\right]\NNbare Q^\bare\,.\label{sigSq}
\eea
so we have the collinear factorizations
\bea
D_{qg}^\bare &=& \Gamma_{qq}^N\left[(D_{qq}+ \smallfrac{1}{2n_f}D_{q\qb})
2n_f\Gamma_{qg}
+D_{qg}\Gamma_{gg} \right],\label{collfactDqg}\\
(D_{qq}^\bare+\smallfrac{1}{2n_f}D_{q\qb}^\bare)&=&
\Gamma_{qq}^N\left[(D_{qq}+ \smallfrac{1}{2n_f}D_{q\qb})\Gamma_{qq}
+D_{qg}\Gamma_{gq}\right].\label{collfactDqq}
\eea

Now consider first the gluonic contribution.
Using eqns.~(\ref{heNS}) and eqns.~(\ref{heS}), then
since $\Gamma_{qq}^N=1+\ord(\as)$, while $\Gamma_{qg}=
\ord(\as(\smallfrac{\as}{N})^k)$
\beq \label{Dqgcol}
D_{qg}^{\bare}= \Gamma_{qg}+D_{qg}\,\Gamma_{gg}
+ \ord \left(\as^2 \left(\smallfrac{\as}{N}\right)^k \right)\,.
\eeq
We would like to factorize out an overall factor of
$\Gamma_{gg}$, in order to compare to the $k_T$-factorization
eqn.~(\ref{xsecmelSkt}). This can be achieved
by taking the logarithmic derivative with respect to $Q^2$ \cite{ch}:
since
\bea
\frac{\partial}{\partial \ln Q^2}
\Gamma_{gg}(N,\as(\smallfrac{Q^2}{\mu^2})^\eps,\eps)
&=& \gamma_{gg}(N,\as(\smallfrac{Q^2}{\mu^2})^\eps S_\eps)
\Gamma_{gg}(N,\as(\smallfrac{Q^2}{\mu^2})^\eps,\eps)\label{derivGgg},\\
\frac{\partial}{\partial \ln Q^2}
\Gamma_{qg}(N,\as(\smallfrac{Q^2}{\mu^2})^\eps,\eps)
&=& \gamma_{qg}(N,\as(\smallfrac{Q^2}{\mu^2})^\eps S_\eps)
\Gamma_{gg}(N,\as(\smallfrac{Q^2}{\mu^2})^\eps,\eps)\label{derivGqg},
\eea
we find
\begin{equation} \label{derDqgcol}
 \frac{\partial}{\partial \ln Q^2}D_{qg}^\bare\Big\vert_{\mu^2=Q^2}
=\left[\gamma_{qg}
+\gamma_{gg}D_{qg}+ \eps\as\frac{\partial}{\partial\as}D_{qg}
\right] \Gamma_{gg}
+ \ord (\as^2 \left(\smallfrac{\as}{N}\right)^k )\,.
\end{equation}
On the other hand, from eqn.~(\ref{dimregqgbare})
the high energy factorization gives
\beq\label{derDqghe}
\frac{\partial}{\partial \ln Q^2} D_{qg}^\bare\Big\vert_{\mu^2=Q^2}
= h_{qg}(N,\gamma_{gg};\as)R(\gamma_{gg})
\Gamma_{gg}(N;\as,\eps)+\ord(\as^2(\smallfrac{\as}{N})^k).
\eeq
Comparing eqns.~\eqn{derDqgcol} and \eqn{derDqghe}, we can factor
all the collinear singularities into $\Gamma_{gg}$ on each side, and thus
we must have
\beq\label{Dqgresult}
\gamma_{qg}(N,\as)+ \gamma_s(\smallfrac{\as}{N})
D_{qg}(N,\as) = h_{qg}(N,\gamma_s;\as)
R(\gamma_s)+\ord(\as^2(\smallfrac{\as}{N})^k)\,,
\eeq
where we also used eqn.~\eqn{ggghe}. We can thus reduce the calculation of
$D_{qg}$ in \MS\ factorization to a calculation of the impact factor
eqn.~\eqn{impfact}
since $\gamma_s$, $\gamma_{qg}$ and $R(\gamma_s)$ are already known to the
required order \cite{ch}.

The quark coefficient function $D_{qq}$ may be determined similarly:
since $D_{q\qb} = 1+\ord(\as)$ and $D_{q\qb}^\bare = 1 +\ord(\as)$,
eqn.\eqn{collfactDqq} may be written
\beq \label{Dqqcol}
D_{qq}^{\bare}= D_{qq}+\Gamma_{qq}^{PS}+D_{qg}\,\Gamma_{gq}
+ \ord (\as^2 \left(\smallfrac{\as}{N}\right)^k )\,.
\eeq
At high energy this is given by eqn.~\eqn{dimregqqbare}. The subtraction
term is just the on-shell partonic cross-section, i.e. the bare
coefficient function evaluated at $\ord(\as)$. It may be evaluated using
eqs.~\eqn{Dqgcol} and \eqn{transqg}:
\beq\label{onshellDqg}
\Sigma_{qg}(N,0;\as,\eps) =
\smallfrac{\as}{\eps}\gamma_{qg}^0(N)+\as D_{qg}^0(N)+\ord(\as^2)
\eeq
where the first term is the collinear singularity from the
quark-gluon splitting and the second the renormalized coefficient function
to $\ord(\as)$. Substituting eqs.~\eqn{Dqgcol}, \eqn{Dqqcol} and
\eqn{onshellDqg} into the bare colour charge relation
eqn.~\eqn{dimregqqbare} we obtain:
\beq\label{ccintermediate}
D_{qq}+\Gamma_{qq}^{PS}+D_{qg}\,\Gamma_{gq}
=  \smallfrac{C_F}{C_A}
 \left[\Gamma_{qg}+D_{qg}\,\Gamma_{gg}
-\smallfrac{\as}{\eps}\gamma_{qg}^0
-\as D_{qg}^0
 \right]+\ord(\as^2\left(\smallfrac{\as}{N}\right)^k) \,.
 \eeq
Using the colour charge relations \eqn{transqq} and \eqn{transgq}
all the singular terms cancel as they should, and we are left with
the renormalized colour charge relation for the Drell-Yan
coefficient functions
\beq\label{quarkcoeff}
D_{qq}\left(N,\as \right)
= \smallfrac{C_F}{C_A} \left[D_{qg}\left(N,\as\right)
- \as D_{qg}^0(0)\right] +
\ord(\as^2\left(\smallfrac{\as}{N}\right)^k)\,.
\eeq
Thus the high energy behaviour of the pure singlet quark coefficient
function $D_{qq}$ may be determined directly from that of the quark
gluon coefficient function $D_{qg}$ found from the
eqn.~(\ref{impfact}) through eqn.~\eqn{Dqgresult}.

Finally we consider the factorization of the remaining contribution to the
Drell-Yan cross-section, $\sigma^{SS}$ given by eqn.~\eqn{sigSS}. The
terms in the first line are $\ord(\as(\smallfrac{\as}{N})^k)$, while those
in the second are $\ord(\as^2(\smallfrac{\as}{N})^k)$, so we consider
them separately. From the coefficients of $Q^\bare G^\bare$ and
$Q^\bare Q^\bare$ in the first line we now find (in the same way that we
derived eqns. \eqn{collfactDqg} and \eqn{collfactDqq})
\bea
D_{qg}^\bare &=& 2(D_{qq}+ \smallfrac{1}{4n_f}D_{q\qb})
\Gamma_{qq}2n_f\Gamma_{qg}
+D_{qg}(\Gamma_{qq}\Gamma_{gg}+2n_f\Gamma_{qg}\Gamma_{gq}),
\label{collfactDqgSS}\\
(D_{qq}^\bare+\smallfrac{1}{4n_f}D_{q\qb}^\bare)&=&
(D_{qq}+ \smallfrac{1}{4n_f}D_{q\qb})\Gamma_{qq}^2
+D_{qg}\Gamma_{qq}\Gamma_{gq}.\label{collfactDqqSS}
\eea
Keeping only terms of $\ord(\as(\smallfrac{\as}{N})^k)$, these simplify to
eqns. \eqn{Dqgcol} and \eqn{Dqqcol}, as they must.

From the second line of eqn.~\eqn{sigSS}
consider the gluon-gluon term: it is easy to show that this may be used
to determine the resummed coefficient function $D_{gg}$ in terms of an
impact factor computed as the double Mellin transform of the
second derivative of the off-shell partonic cross-section
$\Sigma_{gg}(N,\smallfrac{\kt_1^2}{Q^2},\smallfrac{\kt_2^2}{Q^2};\as,\eps)$
in eqn.~\eqn{xsecmelSkt}: the result is
\beq\label{Dggresult}
\gamma_{qg}^2+ \gamma_{qg}\gamma_s
D_{qg}+\gamma_s^2D_{gg} = h_{gg}(N,\gamma_s,\gamma_s;\as)
R(\gamma_s)^2+\ord(\as^2(\smallfrac{\as}{N})^k)\,,
\eeq
where
\beq\label{impfactgg}
h_{gg}(N,M_1,M_2;\as)\equiv
M_1^2 M_2^2
\int_0^\infty\!d\xi_1\, \xi_1^{M-1}
\int_0^\infty\!d\xi_2\, \xi_2^{M_2-1}
{\Sigma}_{gg}(N,\xi_1,\xi_2;\as,0).
\eeq
However in this case the bare colour-charge relations
implicit in the $k_T$-factorization do not lead directly to simple
colour-charge relations among the coefficients $\tilde D_{qq}$,
$\tilde D_{qg}$ and $D_{gg}$, as they did for heavy quark
production \cite{ellis-hq}, because here there are also contributions of
the same order from $\ord(\as^2(\smallfrac{\as}{N})^k)$
terms in $D_{qq}$ and $D_{qg}$, and these cannot at present be computed
until $k_T$-factorization is extended to NNLL$x$.

For the remainder of this paper, we will work entirely at NLL$x$, computing
the impact factor $h_{qg}$ and thus all the $\ord(\as(\smallfrac{\as}{N})^k)$
contributions to the coefficient functions $D_{qq}$ and $D_{qg}$.

\section{The off-shell cross-section}

In this section we compute the off-shell cross-section
$\Sigma_{qg}(\tau,\xi;\as,0)$
for the process $$g^*(k)\; q(p) \to \gamma^* (q) \;q(p')\,,$$ where
\begin{eqnarray} \label{kin}
 k &=& x_1 p_1 + \kt, \qquad p = x_2 p_2,
 \nonumber \\
q &=& z_1 x_1 p_1+ (1-z_2) x_2 p_2  + \qt, \\
 p' &=&  (1-z_1) x_1 p_1+ z_2 x_2 p_2  + \kt- \qt,\nonumber
\end{eqnarray}
and we define the dimensionless variables
\begin{equation}\label{vars}
 \tau = \frac{Q^2}{\nu}, \qquad \xi= \frac{|\kt|^2}{Q^2},
\end{equation}
with $Q^2=q^2$ and $\nu = 2 x_1x_2 p_1\cdot p_2$.
We need to consider two diagrams, given in fig.~\ref{dyqg}.
Since soft and collinear divergences are regulated by the off-shellness of
the incoming gluon, we may perform the whole calculation in four dimensions.
The two-body phase-space is then given by
\begin{eqnarray}\label{phasespace}
 d \Phi^{(2)}& =& \frac{d^4 q}{(2 \pi)^3}
\frac{d^4 p'}{(2 \pi)^3}\,\delta (q^2-Q^2)
\delta (p'^2) (2 \pi)^4 \delta^{(4)} (k+p-q-p')
  \nonumber \\
 &=&\frac{d^4 q}{(2 \pi)^2}\, \delta (q^2-Q^2)\, \delta (p'^2)=\nonumber \\
 &=& \frac{\nu}{8 \pi^2} dz_1 dz_2 d^2 \qt \,
\delta((1-z_2) z_1 \nu - |\qt|^2-Q^2)\delta((1-z_1) z_2 \nu - |\kt-\qt|^2).
\end{eqnarray}
The squared matrix element is computed using the usual eikonal
polarisations for the gluon; the photon indices are contracted
with $g^{\mu \nu}$, because of conservation of electro-magnetic
current, just as in the on-shell calculation \cite{Altarelli:1978id}.
The result is
\bea \label{offshell} |\m|^2 &=& -\frac{e_q^2 g_s^2}{N_c}
\Big\{\frac{t}{s}+\frac{s}{t}
+Q^2|\kt|^2\left(\frac{1}{s^2} +\frac{1}{t^2} \right)+
 + 4\frac{\kt\cdot\qt}{s} -4\frac{Q^2\kt\cdot\qt}{t^2}\left(1-
\frac{\kt\cdot\qt}{|\kt|^2}\right)\nonumber \\ &&
\qquad\qquad + \frac{2}{st}
\Big[(Q^2+|\kt|^2)(|\kt|^2-2\kt\cdot\qt)
+2(\kt \cdot \qt)^2+|\kt^2|(s-t)\Big] \Big \}
 \eea
 where the Mandelstam invariants are
 \bea\label{mandelstam}
s &=& (p+k)^2 = \nu -|\kt|^2, \nonumber \\
t &=& (p-q)^2 = Q^2 - z_1\nu, \nonumber \\
u &=& Q^2- |\kt|^2-s-t. \eea
In the on-shell limit $|\kt|\to 0$, when averaged over
the angle $\vartheta$ between $\kt$ and $\qt$, the off-shell matrix
element eqn.~(\ref{offshell})
reduces to
 \beq \label{onshell}
\lim_{|\kt|\to 0}\langle
|\m|^2\rangle_{\vartheta}= -\frac{e_q^2 g_s^2}{N_c}
\Big\{\frac{t}{s}+\frac{s}{t}+ 2 \frac{Q^2u}{st} \Big\}\,,
 \eeq
in agreement with the standard calculation.

To perform the phase space integration we first
change the variable of integration by introducing $\Delta = \qt -
z_1 \kt$, and we then use one delta function to perform the integral
over $z_2$, and the second to perform the integral over $\Delta^2=
|\qt|^2-2z_1\kt\cdot\qt+z_1^2|\kt|^2$,
i.e. we write
\bea\label{doubledelta}
d \Phi^{(2)}&=& \frac{1}{8 \pi^2}
\frac{dz_1\,dz_2}{(1-z_1)} d^2 \Delta \,\delta((1-z_2) z_1 \nu -
|\qt|^2-Q^2)\delta(z_2-\frac{|\kt-\qt|^2}{(1-z_1) \nu})\nonumber\\
&=& \frac{1}{16 \pi^2} dz_1\, d\vartheta\, dz_2\, d\Delta^2\,
\delta\big(z_2-\smallfrac{|\kt-\qt|^2}{(1-z_1) \nu}\big)
\delta\big(\Delta^2 - (1-z_1)[z_1(\nu-|\kt|^2)-Q^2]\big).
\eea
The remaining integral over $\vartheta$ then runs from zero to $2\pi$,
while the integral over $z_1$ runs from $Q^2/(\nu-|\kt|^2)$ to one.
The dimensionless cross-section can then be written as
\beq\label{twointsleft}
\Sigma_{qg}(\tau, \xi;\as,0)
= \frac{\as}{2 \pi}\frac{\tau}{2}
\int_{\frac{\tau}{1-\tau \xi}}^1 \!d z_1\,\int_0^{2 \pi}
\!\frac{d \vartheta}{2 \pi}\,|\m|^2\big\vert_{z_2,\Delta^2},
\eeq
where $|\m|^2\big\vert_{z_2,\Delta^2}$ is the squared matrix element
\eqn{offshell} with the two $\delta$-functions \eqn{doubledelta} imposed.
The normalization is consistent with eqn.~(\ref{dylo}), with in particular the
factor $\smallfrac{e_q^2}{N_c}$ has been absorbed into the LO cross-section.
The two remaining integrations can now be performed explicitly: the
result is
 \bea \label{DYosxsec}
 \Sigma_{qg}(\tau, \xi;\as,0) &=&\frac{\as}{2 \pi}T_R\,
\tau\Bigg\{\frac{1}{1- \tau \xi}\ln \left(\smallfrac{(1-\tau)
(1- \xi \tau)}{\tau^2 \xi} \right)p_1(\tau,\xi)
 \nonumber \\
 &&\qquad\qquad\qquad\qquad +\frac{1-\tau-\tau \xi}{2(1-\tau)(1-\tau \xi)^3}
p_2(\tau,\xi)\Bigg\}
\Theta\left(\frac{1}{\tau}-\xi-1 \right)\,.
 \eea
where the polynomials
\bea\label{polly}
p_1(\tau,\xi)&=& \tau^2+(1-\tau)^2
+ \tau^2 \xi \left(10+ \xi +18 \tau^2 \xi -6 \tau (3+2 \xi)\right),
\nonumber\\
p_2(\tau,\xi)&=&-1+36 \tau^5 \xi^3+7 \tau (2+\xi)-6 \tau^4 \xi^2(15+7 \xi)
\\&&\qquad\qquad\qquad
+2 \tau^3 \xi (35+49 \xi+4 \xi^2)
-\tau^2 (15+71 \xi +14 \xi^2).\nonumber
\eea
Note that this cross-section has been already averaged over the
azimuthal angle of the incoming gluon, consistently with
eqns.~\eqn{dimregkt} and \eqn{impfact},
where the Mellin integral is taken with respect to $\xi = \kt^2/Q^2$.
The on-shell limit $\xi \to 0$ is
\beq
\Sigma_{qg}(\tau, \xi;\as,0) =
\frac{\as}{2 \pi}\bigg[\tau P_{qg}(\tau)
\ln\big(\smallfrac{1-\tau}{\xi\tau^2} \big)
+ \smallfrac{1}{2}\tau\left(-\smallfrac{1}{2}+ 7 \tau
-\smallfrac{15}{2} \tau^2 \right) \bigg]+ \ord(\xi)\,,
\eeq
 where $P_{qg}(\tau)= T_R(\tau^2+(1-\tau)^2)$ is the LO
$qg$-splitting function.
Because of the off-shell regularisation, this result is not
the same as the usual $\msb$ NLO coefficient function
\cite{Altarelli:1978id}:
\beq\label{usualNLO}
D^{1}_{qg}(\tau;\as,\eps) =
\frac{\as}{2 \pi}\bigg[\tau P_{qg}(\tau) \Big(\frac{1}{\eps}+\ln 4 \pi-
\gamma_E+\ln\big(\smallfrac{(1-\tau)^2}{\tau}\big) \Big)
+ \smallfrac{1}{2}\tau(\smallfrac{1}{2}+ 3 \tau
-\smallfrac{7}{2} \tau^2) \bigg],
\eeq
though the collinear singularities match as they must, with
$\ln(1/\xi) \to 1/\eps +\ln 4 \pi- \gamma_E $.

\section{The quark-gluon impact factor}

The quark-gluon impact factor is defined as the double Mellin transform of the
double logarithmic derivative of the off-shell cross-section (see
eqn.~(\ref{impfact}):
\beq\label{impfactdM}
h_{qg}(N,M;\as)=  M^2\int_0^{\infty} \xi^{M-1}\int_0^1 \tau^{N-1}
\Sigma(\tau,\xi;\as,0)\,.
\eeq
Because the off-shell cross-section eqn.~\eqn{DYosxsec} contains a
factor $\Theta(\tau^{-1}-\xi-1)$, it is useful to change the
integration variables according to
\beq\label{alphabeta}
\alpha = \tau \xi,\qquad
\beta = \frac{\tau}{1-\tau \xi};
\eeq
the Jacobian determinant is $1/\beta$ and the $\Theta$-function
condition is satisfied for all $\alpha,\beta \in [0,1]$, so
\bea\label{impfactab}
h_{qg}(N,M;\as,0)&=& \frac{\as}{2 \pi}T_R\, M^2
\int_0^1 \!d \alpha\, \int_0^1 \! d \beta\,
\alpha^{M-1}(1-\alpha)^{N-M}\beta^{N-M} \nonumber \\
&&\qquad \left[ \ln \smallfrac{1-(1-\alpha) \beta}
{\alpha \beta} d_1(\alpha,\beta)
-\smallfrac{1}{2}\frac{(1-\beta)}{1-(1-\alpha) \beta}
d_2(\alpha,\beta) \right]\,,
\eea
where the polynomials $d_1$ and $d_2$ are now
\bea\label{polyd}
d_1(\alpha,\beta)&=&  1 + \alpha^2  - 2 \beta + 12 \alpha \beta
- 22 \alpha^2  \beta + 12 \alpha^3  \beta + 2 \beta^2
- 22 \alpha \beta^2  \nonumber \\&&\qquad +  56 \alpha^2  \beta^2
- 54 \alpha^3  \beta^2  + 18 \alpha^4  \beta^2\,, \nonumber \\
d_2(\alpha,\beta)&=&  1 -6 \alpha + 8\alpha^2  - 14 \beta
+ 71 \alpha \beta - 98 \alpha^2  \beta + 42 \alpha^3  \beta
+ 15 \beta^2  \nonumber \\&&\qquad - 85 \alpha \beta^2
+  160 \alpha^2  \beta^2  - 126 \alpha^3  \beta^2  + 36 \alpha^4  \beta^2\,.
\eea
The integrals containing the logarithm can be further simplified through
integrating by parts with respect to the variable $\beta$:
\bea \label{parts}
\int_0^1 d \alpha \int_0^1 d \beta\, \alpha^{M-1+p}
(1-\alpha)^{N-M}\beta^{N-M+q}\ln \smallfrac{1-(1-\alpha) \beta}
{\alpha \beta}= \qquad \qquad\qquad\nonumber \\
\frac{1}{1+N-M+q}\int_0^1 d \alpha \int_0^1 d \beta\,
\alpha^{M-1+p}(1-\alpha)^{N-M}\beta^{N-M+q}
\frac{1}{1-(1-\alpha) \beta}.
\eea
where $p$ and $q$ are integers, the exponents of $\alpha$ and
$\beta$ in each terms of the functions $d_1(\alpha,\beta)$.
Thus the only nontrivial integral which we need is
\bea
 &&\int_0^1 d \alpha \int_0^1 d \beta\, \alpha^{M-1+p}
(1-\alpha)^{N-M}\beta^{N-M+q}\frac{1}{1-(1-\alpha) \beta}=
\nonumber \\
&&\qquad =\sum_{k=0}^{\infty}
\int_0^1 d \alpha \int_0^1 d \beta\alpha^{M-1+p+k} (1-\alpha)^{N-M}
\beta^{N-M+q+k} \nonumber \\
&&\qquad =\sum_{k=0}^{\infty}
\frac{\Gamma(1+N-M+k)\Gamma(M+p)}{\Gamma(1+N+k+p)(1+N-M+q+k)} \label{infsum} \\
&&\qquad =
 \frac{\Gamma(M+p)\Gamma(1+N-M)}{\Gamma(1+N+p)(1+N-M+q)}\nonumber \\
 &&\qquad\qquad{}_3F_2(1,1+N-M,1+N-M+q;1+N+p,2+N-M+q;1)\nonumber \\
&&\qquad =
 \Gamma^2(M+p)
\frac{\Gamma(1+N-M)\Gamma(1+N-M+q)}{\Gamma(1+N+p)\Gamma(1+N+p+q)}\nonumber \\
 &&\qquad\qquad{}_3F_2(M+p,N+p,1+N-M+q;1+N+p,1+N+p+q;1),\label{master}
\eea
where ${}_3F_2(a,b,c;d,e;z)$ is a generalised hypergeometric
function, and in the last line we used Thomae's theorem.
The final result for $h_{qg}$ eqn.~\eqn{impfactab} is then a
fairly complicated sum of
terms involving the generalised hypergeometric functions ${}_3F_2$:
\bea \label{fullDYimpact}
h_{qg}(N,M;\as,0)&=& \frac{\as}{2 \pi}T_R \,
\sum_{p=0}^4 \sum_{q=0}^3
M^2\Gamma^2(M+p)
\frac{\Gamma(1+N-M)\Gamma(1+N-M+q)}{\Gamma(1+N+p)\Gamma(1+N+p+q)}\nonumber \\
 &&\qquad{}_3F_2(M+p,N+p,1+N-M+q;1+N+p,1+N+p+q;1)
\nonumber \\
&&\qquad\qquad\qquad\Big[\frac{{\Delta}^{(1)}_{1+p,1+ q}}{(1+N-M+q)}
+{\Delta}^{(2)}_{1+p,1+ q}  \Big]
\eea
where
\begin{equation}
 {\Delta}^{(1)}_{i,j}= \left(
\begin{array}{cccc}
1 & -2 & 2 &0\\
0 & 12 & -22 &0 \\
1 & -22 & 56& 0\\
0 & 12 & -54&0 \\
0 & 0 & 18&0
\end{array}
\right)\,,
\quad
 {\Delta}^{(2)}_{i,j} =-\smallfrac{1}{2}\left(
\begin{array}{cccc}
1 & -15 & 29&-15 \\
-6 & 77 & -156&85 \\
8 & -106 &258 &-160\\
0 & 42 & -168 &126\\
0 & 0 & 36&-36
\end{array}
\right)\,.
\end{equation}
Note that in the collinear limit $M\to 0$ eqn.~\eqn{fullDYimpact} is regular.
In fact since only terms with $p=0$ survive, and
${}_3F_2(0,N,1+N+q;1+N,1+N+q;1)=1$,
\beq \label{fullDYimpactzM}
h_{qg}(N,0;\as,0)= \frac{\as}{2\pi}T_R\frac{N^2+3N+4}{(N+1)(N+2)(N+3)}
= \as\gamma^0_{qg}(N) \,
\eeq
as it must be from eqn.~\eqn{Dqgresult}. This is a highly nontrivial check on
eqn.~\eqn{fullDYimpact}.

In the limit $N\to 0$ (i.e. in the high energy limit), the sum \eqn{infsum}
may be performed more easily, since the generalised hypergeometric
function reduces to a sum of rational functions of $\Gamma$-functions:
\bea\label{masterzN}
&&\int_0^1 d \alpha \int_0^1 d \beta \, \alpha^{M-1+p}
(1-\alpha)^{-M} \beta^{-M+q} \frac{1}{1-(1-\alpha) \beta}  \nonumber \\
&&\qquad = \sum_{k=0}^{\infty} \frac{\Gamma(1-M+k)\Gamma(M+p)}
{\Gamma(1+k+p)(1-M+q+k)} \nonumber \\
&&\qquad =  \Gamma(M+p)
\Big[\frac{\Gamma(M+p)\Gamma(1-M-p)\Gamma(1-M-p+q)}{\Gamma(1+q)}\nonumber\\
&&\qquad\qquad\qquad\qquad\qquad\qquad
-\sum_{j=0}^{p-1} \frac{\Gamma(1-M-p+j)}{(1-M+q-p+j)\, j!} \Big].
\eea
Using this in eqn.~\eqn{impfactab}, the result remarkably simplifies to
\beq\label{impfactzn}
h_{qg}(0,M;\as,0) =
\frac{\as}{2 \pi}T_R \,4\frac{ \Gamma(1-M)^2\Gamma(1+M)^2}{(1-M)(2-M)(3-M)}\,.
\eeq
The same result may be obtained by setting $N=0$ in the generalised
hypergeometric functions in eqn.~\eqn{fullDYimpact}, and simplifying 
the result. The infrared singularity at $M=1$ is a triple pole, as 
expected from the counting of soft and collinear singularities \cite{ball}.

\section{Drell-Yan coefficient functions at high energy}

The Taylor expansion of eqn.~\eqn{impfactzn} about $M=0$ gives
\bea\label{impfactexp}
h_{qg}(0,M;\as,0) &=& \frac{\as}{2 \pi}T_R
\frac{2}{3}\Big[1+ \frac{11}{6} M
+ \left(\frac{85}{36}+\frac{\pi^2}{3} \right) M^2 +
\left(\frac{575}{216}
+\frac{11 \pi^2}{18} \right) M^3 \nonumber \\ &&+
\left(\frac{3661}{1296}+\frac{85 \pi^2}{108}
+\frac{\pi^4}{15}\right) M^4 +\ord \left(M^5 \right)\Big]
\eea
In order to compute the high energy behaviour of the $\msb$ coefficient
function, we need the scheme dependent factor:  
\beq\label{Rexp}
R(M)= 1+ \smallfrac{8}{3}\zeta_3 M^3 -\smallfrac{3}{4} \zeta_4 M^4
+ \ord \left(M^5 \right)\,.
\eeq
The BFKL anomalous dimension is
 \beq\label{gsexp}
 \gamma_s\left(\frac{\as}{N}\right) =  \frac{C_A \as}{\pi N}
+ 2 \zeta_3 \left(\frac{C_A \as}{\pi N}\right)^4 +
 2 \zeta_5 \left(\frac{C_A \as}{\pi N}\right)^6 + \dots
 \eeq
 while quark anomalous dimension in $\msb$ may be found in \cite{ch}:
\bea \label{gqgexp}
\gamma_{qg}(N,\as) &=& \frac{\as}{3 \pi}T_R \Big[
 1+\frac{5}{3} \frac{C_A}{\pi}\frac{\as}{N}
+\frac{14}{9} \left(\frac{C_A}{\pi}\frac{\as}{N}\right)^2
 + \left(\frac{82}{81}+2 \zeta_3\right) \left(\frac{C_A}{\pi}
\frac{\as}{N}\right)^3 \nonumber \\ &+&
  \left(\frac{122}{243}+\frac{25}{6} \zeta_3\right)
\left(\frac{C_A}{\pi}\frac{\as}{N}\right)^4+
 \left(\frac{146}{729}+\frac{14}{3} \zeta_3+2 \zeta_5\right)
\left(\frac{C_A}{\pi}\frac{\as}{N}\right)^5 +\dots  \Big]\,.
\eea
Substituting eqn.~\eqn{gsexp} into eqns.~\eqn{impfactexp} and \eqn{Rexp}, and
then substituting the results along with eqn.~\eqn{gqgexp} into
eqn.~(\ref{Dqgresult}), the leading terms on either side cancel, and the
remaining terms give the NLL$x$ expansion
\bea
 D_{qg}(N,\as)
&=& \frac{\as}{18 \pi} T_R \Big[1
+ \left(\frac{29}{6}
+{2 \pi^2}\right)\frac{C_A}{\pi}\frac{\as}{N}
\nonumber \\
 &&+\left(\frac{1069}{108}
+\frac{11}{3}\pi^2+{4}\zeta_3 \right)
 \left(\frac{C_A}{\pi}\frac{\as}{N}\right)^2\nonumber \\
 &&+ \left(\frac{9031}{648}+\frac{85}{18}\pi^2+\frac{7}{20} \pi^4
+\frac{73}{3}\zeta_3 \right)
\left(\frac{C_A}{\pi}\frac{\as}{N}\right)^3
 +\dots \Big]\,. 
\eea
The coefficients of $\ord \left(\as \right)$ and
$\ord \left(\as^2 \right)$ are in agreement with the
high energy limit of the fixed order NLO \cite{Altarelli:1978id}
and NNLO \cite{Hamberg:1990np,Blumlein:2005im} computations.
This is a very non-trivial check of the procedure. The
$\ord \left(\as^3 \right)$ and subsequent terms are all new results.

The high energy singularities of the quark-quark coefficient function
are now easily deduced using the colour-charge relation eqn.~\eqn{quarkcoeff}:
we find
\bea
 D_{qq}(N,\as)
&=& \frac{\as}{18\pi}T_R\frac{C_F}{\pi}\frac{\as}{N}
\Big[\left(\frac{29}{6}
+{2\pi^2}\right)
\nonumber \\
 &&+\left(\frac{1069}{108}
+\frac{11}{3}\pi^2+{4}\zeta_3 \right)
 \frac{C_A}{\pi}\frac{\as}{N}\nonumber \\
 &&+ \left(\frac{9031}{648}+\frac{85}{18}\pi^2+\frac{7}{20} \pi^4
+\frac{73}{3}\zeta_3 \right)
\left(\frac{C_A}{\pi}\frac{\as}{N}\right)^2
 +\dots \Big]\,. 
\eea
Again this result checks at $\ord \left(\as^2 \right)$ against the NNLO
result of ref.\cite{Hamberg:1990np,Blumlein:2005im}, while the terms from
$\ord \left(\as^3 \right)$ onwards are new results.

\section{Summary}

We have analysed the Drell-Yan (and thus also vector boson production) 
process in the limit of high partonic centre-of-mass energy, along the lines of 
the analysis of DIS presented in \cite{ch}. We showed how to factorise 
simultaneously high energy and collinear singularities from the hard 
coefficient function, computed the quark-gluon impact factor, and thus
deduced the leading high energy singularities in the Drell-Yan 
coefficient functions in $\msb$ scheme to arbitrarily high 
orders in $\as$. Our results agree with the known results at NLO and 
NNLO, while providing new results at N$^3$LO and beyond.

It will now be possible to use these results to perform an all order 
resummation of high energy logarithms in Drell-Yan processes, using the 
techniques developed in \cite{ball} which have already been applied to 
deep inelastic processes \cite{Altarelli:2008aj}. A global analysis
of the effect of high energy resummation on parton distribution functions and
LHC benchmarks is thus now a definite possibility for the near future.

\bigskip\bigskip
{\bf Acknowledgements:} SM would like to thank SUPA for financial support
during most of the period during which this work was done. We were also
supported in part by the Marie Curie Research and Training network
HEPTOOLS under contract MRTN-CT-2006-035505.
\vfill\eject


\begin{thebibliography}{99}
\baselineskip14pt

\bibitem{Altarelli:1978id}
  G.~Altarelli, R.~K.~Ellis and G.~Martinelli,
  Nucl.\ Phys.\  B {\bf 143} (1978) 521
  [Erratum-ibid.\  B {\bf 146} (1978) 544].

\bibitem{Matsuura:1990ba}
T.~Matsuura, R.~Hamberg and W.~L.~van Neerven,
  Nucl.\ Phys.\  B {\bf 345} (1990) 331.

\bibitem{Hamberg:1990np}
  R.~Hamberg, W.~L.~van Neerven and T.~Matsuura,
  Nucl.\ Phys.\  B {\bf 359} (1991) 343
  [Erratum-ibid.\  B {\bf 644} (2002) 403].

\bibitem{Blumlein:2005im}
  J.~Blumlein and V.~Ravindran,
  Nucl.\ Phys.\  B {\bf 716}, 128 (2005),
  [hep-ph/0501178].

\bibitem{Catani:1989ne}
  S.~Catani and L.~Trentadue,
  Nucl.\ Phys.\  B {\bf 327} (1989) 323.

\bibitem{Catani:2003zt}
  S.~Catani, D.~de Florian, M.~Grazzini and P.~Nason,
  JHEP {\bf 0307}, 028 (2003),
  [arXiv:hep-ph/0306211].

\bibitem{moch}
  S.~Moch and A.~Vogt,
  Phys.\ Lett.\  B {\bf 631} (2005) 48,
  [hep-ph/0508265].

\bibitem{magnea}
 E.~Laenen and L.~Magnea,
 Phys.\ Lett.\  B {\bf 632} (2006) 270,
 [hep-ph/0508284].

\bibitem{Altarelli:2008xp}
  G.~Altarelli, R.~D.~Ball and S.~Forte,
  PoS {\bf RADCOR2007} (2007) 028
  [arXiv:0802.0968 [hep-ph]].

\bibitem{CCH-photoprod}
  S.~Catani, M.~Ciafaloni and F.~Hautmann,
  Phys.\ Lett.\  B {\bf 242} (1990) 97;
 S.~Catani, M.~Ciafaloni and F.~Hautmann,
  Nucl.\ Phys.\  B {\bf 366} (1991) 135.

\bibitem{ch}
  S.~Catani and F.~Hautmann,
  Nucl.\ Phys.\  B {\bf 427} (1994) 475.

\bibitem{ball}
  R.~D.~Ball,
  Nucl.\ Phys.\  B {\bf 796} (2008) 137,
  [arXiv:0708.1277 [hep-ph]]. 

\bibitem{Altarelli:2008aj}
  G.~Altarelli, R.~D.~Ball and S.~Forte,
  Nucl.\ Phys.\  B {\bf 799} (2008) 199,
  [arXiv:0802.0032 [hep-ph]].

\bibitem{ellis-hq}
  R.~D.~Ball and R.~K.~Ellis,
  JHEP {\bf 0105} (2001) 053,
  [hep-ph/0101199].

\bibitem{camici-hq}
  G.~Camici and M.~Ciafaloni,
  Nucl.\ Phys.\  B {\bf 496} (1997) 305
  [Erratum-ibid.\  B {\bf 607} (2001) 431].

\bibitem{Hautmann:2002tu}
  F.~Hautmann,
  Phys.\ Lett.\  B {\bf 535} (2002) 159,
  [arXiv:hep-ph/0203140].

\bibitem{Marzani:2008az}
  S.~Marzani, R.~D.~Ball, V.~Del Duca, S.~Forte and A.~Vicini,
  Nucl.\ Phys.\  B {\bf 800} (2008) 127,
 [arXiv:0801.2544 [hep-ph]].

\bibitem{Marzani:2008ih}
S.~Marzani, R.~D.~Ball, V.~Del Duca, S.~Forte and A.~Vicini,
  Nucl.\ Phys.\ Proc.\ Suppl.\  {\bf 186} (2009) 98
  [arXiv:0809.4934 [hep-ph]].


\end{thebibliography}
\end{document}